# Experiments in Social Media

**Author:** Toby Walsh[1]

**Abstract**.

*Social media platforms like Facebook and Twitter permit experiments to be performed at minimal cost on populations of a size that scientists might previously have dreamt about. For instance, one experiment on Facebook involved over 60 million subjects. Such large scale experiments introduce new challenges as even small effects when multiplied by a large population can have a significant impact.*

*Recent revelations about the use of social media to manipulate voting behaviour compound such concerns. It is believed that the psychometric data used by Cambridge Analytica to target US voters was collected by Dr Aleksandr Kogan from Cambridge University using a personality quiz on Facebook. There is a real risk that researchers wanting to collect data and run experiments on social media platforms in the future will face a public backlash that hinders such studies from being conducted. We suggest that stronger safe guards are put in place to help prevent this, and ensure the public retain confidence in scientists using social media for behavioural and other studies.*

**A case study in voter participation**

We know in detail about one very large experiment using social media as it has been written up in the scientific literature (1). The experiment was designed to improve voter participation. It involved all people aged 18 and over in the United States who used Facebook on 2nd November 2010, the day of the mid-term elections. The 61 million such users of Facebook that day were divided into three randomly chosen groups: one shown a message that "Today is Election Day", others the same message and some thumbnails of their friends who had voted saying "I voted", and the third group not shown anything.

Analysis of this experiment suggests that these interventions increased turnout by about 340,000 additional votes (1). This is around 0.5% of the total number of votes cast. The Facebook experiment wasn't designed to change the outcome. It was simply designed to increase voter participation. In particular, there was no bias in the way users were encouraged to vote. Users for the three difference groups were chosen uniformly at random. However, it is very possible that the experiment changed the outcome of some of the elections held that day. We cannot know for sure as we cannot re-run the election without the intervention. But the evidence points to an impact.

Consider the Windsor-Orange 1 District for the Vermont House of Representatives. The 2010 election in this district was decided by a single vote (2). Similarly the outcome of the 2010 election in the Rutland 5-4 District for the Vermont House of Representatives was also decided by a single vote (3). Both elections were won by a female Democrat running against a male Republican candidate. With such close outcomes, the Facebook experiment could have been critical.

---

[1] UNSW Sydney, Data61 and Technical University Berlin. Email: tw@cse.unsw.edu.au

Suppose, for a moment, that Facebook had a younger and more female demographic in Vermont in 2010 than the voting population of Vermont itself. This is not an unreasonable assumption. Facebook appeals best to adult women aged 18-29 (4). Now suppose younger women in Vermont are more likely to vote for a female Democratic candidate than for a male Republican. Again, that is not an unreasonable assumption.

It follows from these two assumptions that increasing voter participation of Facebook users in Vermont would likely have increased the Democratic vote. This might easily have got one or two extra votes for the Democrats. Given the closeness of the results in these elections, this could have changed the outcome from what would have been a Republican victory without Facebook's experiment to the Democratic win that was actually witnessed.

It should not have been a surprise that this could have happened. Thousands of different elections were held on 2nd November 2010, some of which were very likely to be closely contested. Indeed, Vermont was one of the more obvious places to see a close result. The Vermont House of Representative has relatively small electorates making it easier to have a narrow result, and has historically seen closely fought outcomes. In 1977, 1986, and 2016 there were other districts for the Vermont House of Representatives that were also decided by a single vote.

Facebook ran further experiments to increase voter participation in the 2012 US elections. Less is known about these experiments as they weren't written up in a scientific paper. Facebook claimed they again selected voters at random and didn't focus on a particular group (5). But, as argued earlier, Facebook is not a demographically balanced sample of the US electorate. Running an experiment on random Facebook users may again have impacted the 2012 results.

**Ethical approval**

In July 2010, before the mid-term elections, Dr James Fowler of the University of California, San Diego sought and received approval to run this voter participation experiment from his Institutional Review Board (IRB). The documentation he submitted to the IRB (which he has kindly shared) argued for a waiver of consent from the human subjects being experimented upon as "*the study is minimal risk*", "*users grant Facebook the right to conduct studies like these*", "*The sample size … makes consent acquisition extremely costly*" and seeking consent "*might influence the way subjects respond*". The IRB accepted these arguments and waived consent.

The last argument is reasonable. Telling someone you are running an experiment in voter participation might easily change whether they vote. But the first three arguments are more questionable. The documentation submitted to the IRB claimed that "*the study is minimal risk*" as "*the primary concern is loss of confidentiality*". Shouldn't changing the outcome of some of the elections be considered an important risk? As for prior consent, it is true that users signing up to Facebook must consent to terms and conditions that permit experiments like these. But Facebook's terms and conditions are far too generic and open ended to be approved by any IRB.

As to the final part of the argument for a waiver, acquiring consent on Facebook or any other website need not be costly. For instance, we consent to "cookies" all the time with one simple click.

The documentation submitted to the IRB also promised "*subjects will be provided with additional pertinent information after participation*" by means of "*aggregate, unidentified results in scientific journals*", and posts in the Facebook blog, as well as articles in the media. It is hard to know how well informed subjects were after participation. It was, however, one of the most highly cited articles in Nature from 2012, was picked up by 85 news outlets, and was tweeted about over one thousand times.

**Related experiments**

This is not the only recent experiment using social media that has raised ethical concerns. One of the most high profile experiments involved the manipulation of the news feed of 689,003 people on Facebook with the goal of changing their emotional state (6). The authors observe that the effects are small but "*given the massive scale of social networks such as Facebook, even small effects can have large aggregated consequences*". Whilst the news feed experiment was large, it was one hundred times smaller than the midterm voting turnout experiment. It is difficult to compare the outcomes of the two experiments as they were different in design. However, the effects of the news feed experiment were again very small — a change in the number of positive and negative words in people's status updates of less than 0.1%.

An important difference between the two experiment was that the news feed experiment was not approved by any IRB. It is not known what internal ethics approval process was used by Facebook before the news feed experiment was run over one week in January 2012. However, the authors argue that the experiment "*was consistent with Facebook's Data Use Policy, to which all users agree prior to creating an account on Facebook, constituting informed consent for this research*" (6). Another aspect of the news feed experiment different to the midterm voter participation experiment was that it was less obvious to the subjects of the news feed experiment that they were being experimented upon. In the voter participation experiment, there was an explicit message to encourage users to turn out to vote. In the news feed experiment, users saw what appeared to be their normal news feed.

After the news feed experiment had taken place, the IRB at Cornell University was consulted by one of the authors of the study, Jeffrey Hancock who was then a Professor of Information Science and Communication at Cornell. As Hancock only wished to process and analyse data that had already been collected, the IRB at Cornell University concluded that he was not directly engaged in human research and that no approval was required (7). Inda Verma, the Editor in Chief of PNAS who published the study later wrote that it was "*a matter of concern that the collection of the data by Facebook may have involved practices that were not fully consistent with the principles of obtaining informed consent and allowing participants to opt out*" (7).

**Recommendations**

These incidents suggest that scientists may need to be more careful about running experiments on social media platforms, especially if they are to keep the confidence of the public. We make three recommendations to strengthen ethical approval of such research to help ensure such confidence. The three recommendations address societal impact, data privacy and the information shared with subjects after the study.

The first recommendation is that we may need to take into account not just the impact on the individual under study but the broader impact any experiment might have on society. For a study on voting, this might be an electoral risk. For a study on fake news, it might be decreasing trust within society in real news. For a study on manipulating people's emotions, it might be the emotional wellbeing of the population studied.

The second recommendation is that ethics approval may be needed before receiving or handing over data from an experiment run in the past. You may, for instance, need to obtain informed consent to hand over date to a third party even though there may be no "additional" cost to the individual to pass on their data once it has been collected. Similarly, you may not be able to receive data from someone else without seeking informed consent for this new use of the data. The only circumstances that informed consent may not be needed when handing over or receiving data is when an appropriate IRB gives a waiver. If such a recommendation had been in place, the IRB at Cornell would not have been able to conclude that approval was unnecessary for the analysis of the already collected data in the Facebook news feed experiment.

The third recommendation is that subjects of any experiment may need to be informed directly after the study about the results and their participation. This is especially important when informed consent has been waived and the participants were not aware that they had been the subject of an experiment. It might not be adequate simply to publish the results in the scientific literature or even the wider media. Subjects might have to be contacted directly.

Had such recommendations been in force in 2010, the IRB at the University of California, San Diego might still have granted a waiver to consent for the Facebook experiment on voter participation. However, the IRB might have required a demographical balance to minimise the electoral impact, and might have required Facebook to email every participant after the election with a summary of the outcome. Had these recommendations been in force in 2012, the IRB at Cornell might still have granted a waiver to consent for the Facebook experiment on news feed manipulation. However, they might have required informed consent from the participants, as well as a follow up after the experiment to check on their emotional well being. And had these recommendations been in force in 2014, and Dr Aleksandr Kogan had applied for ethical approval to collect psychometric data in an academic study, he might have been prohibited from sharing this data with Cambridge Analytica.

These recommendations are not the final word on how ethical approval needs to take account of new technologies like social media. It may, for instance, also be desirable if companies running experiments on social media go through a similar ethics approval process as public institutions.

**Acknowledgments:** The author acknowledges support from the European Research Council (AMPLify Advanced Grant 670077) and the Asian Office of Aerospace Research & Development (Grant FA2386-15-1-4016). The author thanks Dr James Fowler for providing copies of the documentation submitted to the IRB at UCSD for approval.

**Bio:** Toby Walsh is Guest Professor at TU Berlin, Scientia Professor of Artificial Intelligence at the University of New South Wales and Data61. He has a B.A. from the University of Cambridge and a M.Sc. and Ph.D. degree from the University of Edinburgh. He is a Fellow of the Australia Academy of Science and of the Association for the Advancement of Artificial Intelligence, a Humboldt Award winner and recipient of the NSW Premier's Prize for Excellence in Engineering and ICT.